\documentclass[onecolumn,prd,superscriptaddress,nofootinbib,notitlepage]{revtex4-1}
\usepackage{amsfonts,amsmath,amssymb,mathrsfs}
\usepackage{hyperref}
\usepackage{color}
\usepackage{graphicx}  % needed for figures
\usepackage{dcolumn}   % needed for some tables
\usepackage{bm}        % for math
\usepackage[english]{babel}

\def\a{\alpha}

\def\r{\rho}
\def\s{\sigma}
\def\t{\tau}
\def\m{\mu}
\def\n{\nu}
\def\k{\kappa}
\def\th{\theta}
\def\g{\gamma}\def\G{\Gamma}
\def\L{t}\def\l{V}
\def\D{\Delta}
\def\la{\langle}
\def\ra{\rangle}
\def\o{\omega}\def\O{\Omega}
\def\d{\delta}
\def\p{\partial}

\def\oxthree{{\cal O}(x^3) }

\def\half{\textstyle{\frac{1}{2}}}

\def\bdoc{\begin{document}}
\def\edoc{\end{document}}
\def\bea{\begin{equation}}
\def\eea{\end{equation}}

\def\beq{\begin{eqnarray}}
\def\eeq{\end{eqnarray}}
\def\be{\begin{eqnarray}}
\def\ee{\end{eqnarray}}
\def\ben{\begin{enumerate}}
\def\een{\end{enumerate}}
\def\la{\langle}
\def\ra{\rangle}
\def\a{\alpha}
\def\g{\gamma}\def\G{\Gamma}
\def\d{\delta}\def\D{\Delta}
\def\e{\epsilon}
\def\z{\zeta}

\def\th{\theta}
\def\k{\kappa}
\def\l{t}
\def\m{\mu}
\def\n{\nu}
\def\o{\omega}
\def\p{\pi}
\def\r{\rho}
\def\s{\sigma}
\def\t{\tau}
\def\L{{\cal L}}
\def\S{\Sigma }
\def\gsim{\; \raisebox{-.8ex}{$\stackrel{\textstyle >}{\sim}$}\;}
\def\lsim{\; \raisebox{-.8ex}{$\stackrel{\textstyle <}{\sim}$}\;}
\def\gtrsim{\gsim}
\def\lessim{\lsim}
\def\loc{{\rm local}}
\def\vm{v_{\rm max}}
\def\bh{\bar{h}}
\def\del{\partial}
\def\nab{\nabla}
\def\half{{\textstyle{\frac{1}{2}}}}
\def\fourth{{\textstyle{\frac{1}{4}}}}

\def\bD{{\bf D}}
\def\bE{{\bf E}}
\def\bF{{\bf F}}
\def\bB{{\bf B}}
\def\bP{{\bf P}}
\def\bV{{\bf v}}
\def\bv{{\bf v}}
\def\bx{{\bf x}}
\def\by{{\bf y}}
\def\bz{{\bf z}}
\def\ba{{\bf a}}
\def\bd{{\bf d}}
\def\bs{{\bf s}}
\def\bn{{\bf n}}
\def\bp{{\bf p}}

\def\O{\Omega}

\def\br{{\bf r}}
\def\bnab{{\bf \nab}}

\def\tE{\tilde{E}}
\def\tL{\tilde{L}}
\def\Horava{Ho\v{r}ava }

\def\oxtwo{\mathscr{O}\left(x^2\right)}
\def\oxthree{\mathscr{O}\left(x^3\right)}
\def\oxfour{\mathscr{O}\left(x^4\right)}
\def\oxfive{\mathscr{O}\left(x^5\right)}
\def\LL{Lanczos-Lovelock}

\def\ph{\phantom}

\begin{document}
\title{ Higher curvature self-interaction corrections to Hawking
  Radiation } \author{C. Fairoos} \email{fairoos.c@iitgn.ac.in}\author{
  Sudipta Sarkar}\email{sudiptas@iitgn.ac.in} \affiliation{Indian
  Institute of Technology, Gandhinagar, 382355, Gujarat , India.}
\author{ K. P. Yogendran \footnote{On leave from Indian
    Institute of Science Education and Research, Mohali} }
\email{pattag@gmail.com} \affiliation{Indian Institute of Science
  Education and Research, Tirupati, 517507, India}

%\date

 \begin{abstract}
The purely thermal nature of Hawking radiation from evaporating black
holes leads to the information loss paradox. A possible route to its
resolution could be if (enough) correlations are shown to be present
in the radiation emitted from evaporating black holes. A re-analysis
of Hawking's derivation including the effects of self-interactions in
GR shows that the emitted radiation does deviate from pure thermality,
however no correlations exist between successively emitted Hawking
quanta. We extend the calculations to Einstein-Gauss-Bonnet gravity
and investigate if higher curvature corrections to the action lead to
some new correlations in the Hawking spectra. The effective trajectory
of a massless shell is determined by solving the constraint equations
and the semiclassical tunneling probability is calculated. As in the
case of general relativity, the radiation is no longer thermal but there is no correlation between successive emissions. The absence of
any extra correlations in the emitted radiations even in Gauss-Bonnet
gravity suggests that the resolution of the paradox is beyond the scope
of semi-classical gravity.
\end{abstract}
\maketitle

\section{Introduction}

Hawking radiation is an intriguing feature of quantum field theory in
curved space-time. Hawking's calculations predict a pure thermal
spectrum from black holes suggesting that collapsing matter in an
initial pure state evolves non-unitarily into a thermal mixed
state. This leads to the information loss paradox in the
semi-classical description of black holes.  A full resolution of the
paradox remains elusive although there are several plausible
mechanisms that salvage unitarity. One possibility is to identify
small deviations from thermality in the Hawking spectrum that arise
from correlations between successive Hawking quanta which could encode
information about the internal structure of the black hole. The usual
derivation of Hawking radiation assumes a fixed space-time background
and therefore neglects back reaction of the emitted radiation on the
space-time metric. Note that a consistent treatment of back reaction
is necessary for the conservation of energy to be valid. In principle,
the back reaction could also induce other correlations in the Hawking
spectrum causing deviations from the pure black body form and opening
up the possibility that the final state is rendered pure again.

The back-reaction problem in its full generality is perhaps a
difficult problem because of the non-linearity of the field
equations. A possible way to extract the essential physics of the
problem is to make use of the simplicity arising from considering
spherically symmetric configurations, i.e., s-wave emission.  This
approach was developed by \cite{Kraus:1994by} who considered a
spherically symmetric shell of radiation propagating in a spherically
symmetric black hole space-time.  Eliminating the gravitational
constraints subject to spherical symmetry allowed them to find an
effective action for the shell (radiation) degrees of freedom.
Surprisingly, the trajectory of the shell, in the massless limit
turned out to be interpretable as a null geodesic of a black hole
geometry with a shifted ADM mass. This simple but remarkable result
was used to quantise the shell action in a WKB approximation. An
explicit calculation of the Bogoliubov coefficients showed that the
emitted radiation deviated from a pure blackbody spectrum.

Subsequently, the calculation of \cite{Kraus:1994by} was recast in
\cite{Parikh:1999mf} as the computation of a semiclassical tunneling
probability for a (spherical) shell of energy $E$ from behind the
horizon (see also \cite{Massar:1998dj, Massar:1999wg}). Once the
effective trajectory of the shell is determined, the tunneling
probability calculated by evaluating the imaginary part of the
semi-classical action is found to be,
\bea
\Gamma (E) \propto e^{ \Delta S}, \label{tunneling}
\eea
where $\Delta S = S(M) - S(M - E)$, the change in the
Hawking-Bekenstein entropy because of the emission of the shell. This
factor is the change of the phase space volume before and after the
emission of the Hawking radiation and is exactly what is expected if
the underlying microscopic theory is unitary. Obviously, the
semi-classical approach can not determine proportionality factor which
depends on the matrix element between the initial and the final state
of the black hole and therefore requires a full quantum theory of
gravity.

The emission probability in Eq. (\ref{tunneling}) contains corrections
due to the self-gravity of the shell. But, because of the exponential
nature of the function, it is immediately clear that,
\bea
\Gamma (E_1) + \Gamma(E_2) = \Gamma(E_1 + E_2),
\eea
where $E_1$ is the energy of a shell tunneling from the hole of
initial mass $M$ and $E_2$ is that of a subsequent shell tunneling out
of the black hole whose mass is $M - E_1$ \cite{Parikh:2004rh}. Hence,
although the emission spectrum is corrected by the self-gravitational
effect of the shell, successive shells tunneling out of the black hole
are not correlated. As a result, the hope of resolving information
paradox using back reaction calculations seems impossible as long as
we are within the semi-classical regime. This result remains true for
the case of a charged shell \cite{Kraus:1994fj} and also for the case
of AdS boundary conditions \cite{Hemming:2000as}.

%% The simple model of a
%% self-gravitating shell seems to be not very useful to resolve the
%% paradox and we continue to believe that the consideration of the back
%% reaction may lead to useful correlation in the Hawking spectrum, we
%% need to look for the possible generalisation of the calculation of
%% \cite{Kraus:1994by}. \\

However, because general relativity is a perturbatively
non-renormalizable theory - it only makes sense as an effective theory
with the Lagrangian written as a series of irrelevant higher curvature
terms. This will, of course, change the dynamics of the shell plus
gravity system raising the question of whether such higher derivative
terms could lead to the correlated emission of Hawking radiation.  In
fact, from the point of view of an effective theory, all terms
consistent with diffeomorphism invariance could be present in the
effective Lagrangian. But, only a subset of such terms may provide a
consistent low energy description. For example, consider the most
general second order higher curvature theory of gravity in $D$
dimensions. The action of such a theory can be expressed as,

\bea
    {\cal A^{G}} = \frac{1}{16 \pi} \int d^D x \sqrt{-g}\left( R
    +\,\alpha\, R^2 +\,\beta \, R_{ab}R^{ab} +
    \, \gamma \, R_{abcd}R^{abcd} \right). \label{Flagg}
\eea

Here the coefficients $ \alpha, \beta, \gamma $ are constants which
measure the departure from general relativity. For generic values of
these coefficients, these higher curvature terms introduce problematic
features in the classical theory. The field equations contain time
derivatives higher than second order creating difficulties for a
well-defined initial value formalism. Moreover, the constraint
structure of such a theory could be different from general relativity
with the addition of possible second class constraints.  This could
lead to ghosts for quantum perturbations around the flat space. Also,
there is no guarantee of the existence of a black hole solution of
arbitrary values of these coefficients. However, for particular values
of these constants, the theory is free of these unpleasantries. In
five dimensions, one such choice is the Einstein-Gauss-Bonnet (EGB)
gravity described by the action,

\bea 
    {\cal A^{G}} = \frac{1}{16 \pi} \int d^5x \sqrt{-g}\left[R
      +\,\alpha\, \left( R^2 -\, 4 \, R_{ab}R^{ab} +\,  R_{abcd} R^{abcd}
      \right)\right]. \label{Flag}
\eea

Einstein-Gauss-Bonnet gravity is the unique theory in five dimensions
which is free from perturbative ghosts \cite{Zwiebach:1985uq} and
leads to a well-defined initial value formalism. The Gauss-Bonnet
correction term also appears as a low energy $\alpha '$ correction in
case of heterotic string theory \cite{Zwiebach:1985uq, Sen:2007qy}.
As in GR, the EGB theory admits spherically symmetric vacuum black
hole solutions of the form \cite{Boulware:1985wk},

\bea
ds^2 = - f(r) dt^2 + \frac{dr^2}{f (r)} + r^2\, d\Omega^2,
\eea

where the metric function is given by,

\bea
f(r)=1+\frac{r^2}{4\alpha}\Big[1-\sqrt{1+\frac{8\alpha M}{r^4}}\Big]. \label{BD}
\eea

Such black holes are analog of asymptotically flat, Schwarzschild
black holes in general relativity and the zero of the metric function
$f(r)$ determines the location of the horizon. It is also possible to
formulate the first law for these black holes for small perturbations
and it can be expressed as \cite{Wald:1993nt, Iyer:1994ys},

\bea
\left(\frac{\kappa}{2 \pi}\right) \delta S = \delta M, \label{1stlaw}
\eea

where $\delta M$ represents the variation of the ADM mass and $\kappa$
is the surface gravity of the hole, if we identify $\kappa / 2 \pi$ as
the Hawking temperature associated with the horizon, $\delta S$
represents the change of the black hole entropy where the entropy $S$
is,

\bea
S = \frac{1}{4}\int_{B} \left( 1 + 2\, \alpha \,{}^{(3)} R\right) \,dA.
\label{JMent}
\eea

The entropy of Einstein-Gauss-Bonnet (EGB) black holes are not
proportional to the area but contain correction terms proportional to
the intrinsic curvature ${}^{(3)} R$ of the three-dimensional
cross-section $B$ of the horizon. This entropy also obeys the second
law for linearized perturbations around a stationary black hole
solution \cite{Sarkar:2013swa}.\\

The aim of this paper is to revisit the original calculation of
\cite{Kraus:1994by} by including a Gauss-Bonnet term in the
gravitational action. We consider only the five-dimensional case,
although the generalization to higher dimensions appears
straightforward. We investigate if the tunneling probability still has
the form given by Eq.(\ref{tunneling}) where $S$ is the appropriate
choice of black hole entropy in Einstein-Gauss-Bonnet gravity, namely
the Jacobson Myers entropy in Eq. (\ref{JMent}) \cite{Sarkar:2013swa}.
The motivation of this study is to understand whether the modification
of the gravitational dynamics due to the introduction of a Gauss
Bonnet term can lead to new correlations in the black hole radiation
spectrum. In fact, we will show that higher curvatures terms, at least
in the form of Gauss-Bonnet gravity do not introduce any new
correlations in the tunneling probability and the resolution of the
paradox requires new physics. \\

To calculate the correction of the Hawking spectrum, we first consider
the Hamiltonian formulation of EGB gravity. We follow $( -, +, +, +,
+)$ signature and our sign conventions are that of \cite{Wald:1984rg}.

\section{Review of Hamiltonian formulation of Einstein-Gauss Bonnet Gravity}

We start with the standard spherically symmetric ADM form of the
metric in five dimensions,

\begin{equation}
  ds^2 = -\left({N^t}\right)^2 dt^2 + L^2\left(dr + N^r dt \right)^2
  + R^2 \Big\{  {d\theta}^2 + \sin^2\theta\left( {d \phi}^2
  + \sin^2\phi\, {d\chi}^2\right) \Big\}. \label{adm}
\end{equation}

The action of a massive shell propagating in this background can be written as,

\begin{equation}
  {\cal A}^{s} = -m\int dt\sqrt{\hat{{N}^t}^2 - \hat{L}^2\left(\dot{\hat{r}}
    +\hat{N}^r\right)^2},
\end{equation}

where the caret over various quantities indicates that those are
evaluated on the world line of the shell.  When the background
geometry is fixed, the trajectory of the shell is simply a time-like
geodesic of the background spacetime. To understand the effects of
self-interaction, we need to solve the field equations in the presence
of the shell. The restriction to spherical symmetry allows us to solve
this problem because the dynamical degrees of freedom of the
gravitational field decouple. The 'Coulomb' part of the gravitational
field is completely fixed by the constraints in the presence of the
shell. Thus, we may 'integrate out' the gravitational fields and
obtain an effective Lagrangian for the shell alone. Note that, since
EGB field equations are also second order in time, the dynamical
degrees of freedom of EGB gravity are same as general
relativity. Also, the structure and nature of the constraints are
similar to GR. Moreover, Birkhoff's theorem holds equally good for
spherically symmetric space-times in EGB gravity \cite{Zegers:2005vx,
  Deser:2005gr}. Hence, it may be expected that the analysis of Kraus
and Wilczek \cite{Kraus:1994by} can be generalized for EGB gravity in
a straightforward manner. Our analysis shows that is indeed the case
except with a minor modification related to the discontinuity
conditions at the location of the shell.\\

To begin, we evaluate the EGB action for the ADM form of the
metric. Spherical symmetry allows us to integrate over the angular
directions and ignoring surface terms, the action takes the form,
\begin{equation}
{\cal A^{G}}  = \int dt \int dr\, \mathcal{L},
\end{equation}

where the Lagrangian is given by \cite{Louko:1996jd},

\beq
\mathcal{L} &=& -\frac{\left[ \dot{L} - (N^rL)' \right]
  \left(\dot{R} - N^rR'\right)}{N^t}
\Bigg\{ R^2 + \lambda \left[1 - \left(\frac{R'}{L}\right)^2
  + \frac{(\dot{R} - N^rR')^2}{3{N^t}^2}\right]\Bigg\}\\ \nonumber
&-& \frac{(\dot{R} - N^rR')^2}{N^t}\left[L R - \lambda \left(\frac{R'}{L}\right)'\right] + N^t L R \left[1 - \left(\frac{R'}{L}\right)^2 \right]
- N^t\left(\frac{R'}{L}\right)'\Bigg\{R^2
+ \lambda \left[1 - \left(\frac{R'}{L}\right)^2 \right] \Bigg\},
\eeq
where, we have redefined the Gauss-Bonnet coupling constant as,
$\lambda = 4 \alpha$. We shall denote $b(L)= \left[1 -
  \left(\frac{R'}{L}\right)^2 \right] $ in what follows.  The
conjugate momenta corresponding to metric variables $L$ and $R$
respectively are,

\begin{equation}
  \pi_L = \frac{\partial \mathcal{L}}{\partial \dot{L}} =
  -\frac{\lambda}{3} y^3 - y\Bigg\{ R^2 + \lambda \,b(L)\Bigg\},
  \label{pilandy}
\end{equation}

and,

\begin{equation}
  \pi_R = \frac{\partial \mathcal{L}}{\partial \dot{R}} =
  -\frac{\left(\dot{L} - (N^rL)'\right)}{N^t} \Bigg\{\lambda \, y^2
  + R^2 +\lambda\,b(L)\Bigg\}- 2 y\Bigg[ LR -\lambda  \left(\frac{R'}{L}
    \right)'\Bigg],
\label{pirandy}
\end{equation}

where we have defined $ y(t) = (\dot{R} - N^rR') / N^t$.  In the GR
limit (i.e. when $\lambda \to 0$), the conjugate momentum $\pi_L=-y
R^2$. In EGB gravity, the relationship is given by Eq.(\ref{pilandy})
which is a cubic equation showing the possibility of many branches of
solutions, of which some may not even have a smooth GR limit.  We can
rewrite the Lagrangian using the variable $y$ as,

\beq
\mathcal{L} &=& -y \left[ \dot{L} - (N^rL)' \right]\Bigg\{R^2 + \lambda\, b(L)+\frac{\lambda \, y^2}{3}\Bigg\} %- \frac{\lambda y^3}{3}\left[ \dot{L} - (N^rL)' \right] 
%- N^t y^2\left[ LR -\lambda  \left(\frac{R'}{L}\right)'\right] \\ \nonumber
+ N^t LR\left(b(L)-y^2\right) 
 - N^t\left(\frac{R'}{L}\right)'\Bigg\{R^2 + \lambda\,b(L)-\lambda y^2 \Bigg\}.
 \eeq

Rearranging the above expression and discarding boundary terms, the
action can be expressed in the standard Hamiltonian form,
 
  \begin{equation}
    {\cal A^{G}}  = \int dt \int dr \Bigg\{ \pi_L \, \dot{L} + \pi_R \,\dot{R}
    - N^t \, {\cal H}_{t}^{G} - N^r \, {\cal H}^{G}_{r} \Bigg \},
 \end{equation}
 
 where %the corresponding Hamiltonian and Momentum constraints are: 
 
 \begin{equation}
\mathcal{H}_t^G = y \pi_R - LR\, \left(b(L)-y^2\right)
 + \Bigg(\frac{R'}{L}\Bigg)'\Bigg\{R^2 + \lambda\, b(L)-\lambda y^2\Bigg\},
 \hspace{1cm} {\rm and} \hspace{1cm}
\mathcal{H}^G_r = R'\, \pi_R - L\, \pi_L' .\label{cons}
 \end{equation}
 
 Note that the quantity $y$ is regarded as a function of the conjugate
 momentum $\pi_L$ as in Eq. (\ref{pilandy}). Thus, the total action of
 the shell and gravity system is,

\begin{equation}
  {\cal A}  = \int dt \, p\, \dot{\hat{r}}
  + \int dt \, dr \Bigg[ \pi_R \dot{R} + \pi_L \dot{L}
    - N^t\left(\mathcal{H}_t^s + \mathcal{H}_t^G \right)
    - N^r\left(\mathcal{H}_r^s + \mathcal{H}^G_r \right)\Bigg]
  - \int dt \, \mathcal{M}_{ADM}, \label{fullaction}
 \end{equation}
  with
 \begin{equation}
 \mathcal{H}_{t}^{s} = \left(\sqrt{\left(\frac{p}{\hat{L}}\right)^2 + m^2}\right) \delta(r-\hat{r}) \quad; \quad \mathcal{H}_r^s = -p\, \delta(r-\hat{r}),
 \end{equation}
 
 and $\mathcal{H}_t^G $ and $\mathcal{H}^G_r $ are given by
 Eq. (\ref{cons}). The full constraint equations are then given by:
 
 \bea
 {\cal H}_t = \mathcal{H}_{t}^{s} + \mathcal{H}_t^G = 0 ; \,\,\, {\cal H}_r= \mathcal{H}_{r}^{s} + \mathcal{H}_r^G = 0.
 \eea 
 
The last term in Eq.(\ref{fullaction}) represents the ADM mass of the
total system which is a functional of the metric variables and needs
to be included for a well-defined variational principle. To obtain an
expression for $\mathcal{M} $, following \cite{Kraus:1994by}, we
consider a linear combination of constraints

\begin{equation}
\frac{R'}{L} \mathcal{H}_t - \frac{y}{L}\mathcal{H}_r = 0.
\end{equation}

Using Eq.(\ref{pilandy}), away from the shell, this can be written as $\mathcal{M'} = 0$, where

\begin{equation}
\mathcal{M} = \frac{1}{2}\Bigg(y^2 +b(L)\Bigg)R^2+\frac{\lambda}{4}\Bigg(y^2 + b(L)\Bigg)^2. \label{mass}
\end{equation}

Therefore, ${\cal M}$ is a constant away from the shell and the shell
causes a discontinuity in the value of ${\cal M}$.  To understand this
better, consider a static slice ($ \pi_L = 0$). We can use
Eq.(\ref{pilandy}) to solve for $y$ and it turns out the only solution
which has a smooth GR limit is $y = 0$. Then, using a Schwarzschild
type gauge condition $ R' =1 $ and $ R=r$, we can then solve for the
metric variable $L$ from Eq.(\ref{mass}) and obtain,
 
 \begin{equation}
   \frac{1}{L^2} = 1 + \frac{r^2}{\lambda}
   \Bigg[1 - \sqrt{1+\frac{4\lambda \mathcal{M}}{r^4}}\Bigg].
 \end{equation}
 
This exactly matches with the static slice of the spherically
symmetric vacuum solution of EGB gravity in Eq.(\ref{BD}) with ${\cal
  M}$ as the corresponding mass parameter. Therefore outside the shell
(i.e. $ r > \hat{r})$, we have ${\cal M} = M_{+}$, the ADM mass of the
total shell-gravity system. Inside the shell (i.e. $ r < \hat{r})$,
let us denote ${\cal M} = M$. The relationship between $M$ and $M_{+}$
is obtained by solving the constraints at the position of the shell.

\section{Effective action of the shell}

We shall now follow a procedure similar to \cite{Kraus:1994by}, and
integrate out the gravitational degrees of freedom to find the
equation of motion for the shell. The guiding principle is that, after
elimination of the Lagrange multipliers $N^t, N^r$, under a variation
of the fields subject to the lapse and shift constraints, the
variation of the effective action over all spacetime should take the
Hamilton-Jacobi form,

\begin{equation}
  \delta {\cal A} = \int dt \, p_C\, \dot{\hat{r}} + \int dt \, dr
  \left( \pi_R \delta{R} + \pi_L \delta{L} \right)- \int dt \,
 {M}_{+} .\label{fullactionv}
\end{equation}

In this phase space of configurations (away from the shell), $\pi_L$
and $\pi_R$ have already been determined in terms of the $L$ and $R$
as in Eq. (\ref{pilandy}) and Eq.(\ref{pirandy}) respectively.
Therefore, we can find ${\cal A}$ by substituting for $\pi_{L,R}$ in
the above equation and integrating over field space.

We will integrate along a contour in $L$-space and $R$-space (over
which the constraints are satisfied) starting from the given geometry
specified by $L, R, \pi_{L},\pi_R$ up to some fiducial geometry which
has $\pi_{R, L}=0$. Here we implicitly assume that $L, \, R$
configuration sub-space is homotopically trivial (otherwise the choice
of the contour of integration will matter). This is done in two
stages- integrate $\int \pi_L dL$ keeping $R$ fixed with $\pi_L$
determined by Equation(\ref{pilandy}) until $\pi_L=0$. Subsequently,
we integrate along the $\pi_L=0$ contour (varying both $L$ and $R$) to
the fiducial geometry.

Before proceeding to perform the integrals, we rewrite Eq. (\ref{pilandy}) as,
\bea \pi_L=-\left[\frac{\lambda
    \, y^3}{3}+\lambda\, y\, \left(a(R)-y^2\right)\right],
\eea
where  $a(R)=\sqrt{\frac{4\mathcal{M}}{\lambda}+\frac{R^4}{\lambda^2}}$.

Here we have used the definition Eq. (\ref{mass}), of ADM mass which
can also be written as
\begin{equation}
  \left(b(L)+y^2+\frac{R^2}{\lambda}\right)^2=\left(\frac{4\mathcal{M}}{\lambda}+\frac{R^4}{\lambda^2}\right).
  \label{constraint}
\end{equation}
From this, assuming that $R$ is being held fixed we get $dL=-(yL^3 /
R'^2) \,dy$ where we consider $ L = L(y)$. This allows us to change
the variable of integration from $L$ to the field $y$.  At the lower
limit of integration, the value of $y$ is determined by requiring that
$\pi_L=0$ - which is a cubic equation for $y$. But we have already
observed that $y=0$ gives the GR limit. Therefore, we integrate along
the real $y$-contour until $y=0$.

Thus we get
\begin{equation}
  {\cal A} = - \int_{r_{min}}^{\infty} dr \int_0 ^y \frac{R' \pi_L(y)\, y\,
    dy}{(k^2+y^2)^{\frac{3}{2}}},
\end{equation}
where we have defined
$k^2=1-a(R)+\left(R^2 / \lambda\right).$

The integral can be done exactly and we get 
%% \begin{equation}
%% S=-\int dr \frac{R'}{3} \left(\frac{\lambda y \left(3 a+3
%%   k^2+y^2\right)}{\sqrt{k^2+y^2}}-3 \lambda \left(a+k^2\right) \log
%% \left(\sqrt{k^2+y^2}+y\right)\right)
%% \end{equation}
%% %In this expression $y$ is regarded as a function $y(L,R,R')$, and 
%% %Upon using $k^2+a=1+\frac{R^2}{\lambda}$ and $k^2+y^2=\frac{R'^2}{L^2}$ and
%% which simplifies amazingly to give 
\begin{equation}
 {\cal A} =-\int dr \left(L\, y\left(\lambda+R^2+\frac{\lambda y^2}{3}\right) -
  R'\left(\lambda+R^2 \right)
  \log\left(\frac{\frac{R'}{L}+y}{k}\right) \right),
   \label{Saction}
\end{equation}
where the variable $y$ is an implicit function $y(L,R,R')$ as
determined from the constraint Eq. (\ref{mass}). The above derivation
works away from the shell, and hence the coordinate $r$ is to be
integrated over all space excluding the location of the shell at $r=
\hat r$.\\

We can complete the evaluation of the action by integrating along the
$\pi_L=0$ contour by varying $R$ until we attain the fiducial
metric. But since $\pi_L=0$, the second constraint $\pi_R=(L / R')
\pi_L '$ implies that $\pi_R=0$ - hence, this integral doesn't
contribute anything to the action as in \cite{Kraus:1994by}. It is
obvious that the $\lambda\to 0$ limit reduces to the GR action (the
powers of $R$ are different from \cite{Kraus:1994by} since we are in
five dimensions).

By construction, the action in Eq. (\ref{Saction}) satisfies
$\pi_L=\left(\delta {\cal A} / \delta L \right)$. But since
derivatives of $R$ appear in the action, an integration by parts is
required in determining $\left( \delta {\cal A} / \delta
R\right)=\pi_R$. Remarkably, one can show that the second constraint
$\pi_R =\left(L / R' \right) \pi_L'$ holds automatically, modulo terms
at the location of the shell which spoil this identity. These surface
terms at $r=\hat r$ are similar to the ones in \cite{Kraus:1994by},
\begin{equation}
  \left[\left(\frac{\partial  {\cal A} }{\partial R'}\right)(\hat{r}+\epsilon)
    - \left(\frac{\partial  {\cal A} }{\partial R'}\right)
    (\hat{r}-\epsilon)\right] dR .\label{compensation}
\end{equation}
We can evaluate these terms, $\left( \partial {\cal A} / \partial R'
\right) $ and subtract them from the action which will ensure that
$\pi_R=\left(\delta {\cal A} / \delta R \right)$ everywhere.

Similarly, under an arbitrary variation $\d R$, the ADM mass term in
the action changes as well because the $y-$field implicitly depends on
$R$. Including these surface terms, we arrive at
\begin{equation}
  \frac{\partial {\cal A}}{\partial M_+} d{M_+}
  = \int _{\hat{r}+\epsilon}^{\infty} dr
  \frac{L y}{\left(\left(\frac{R'}{L}\right)^2 - y^2\right)} d{M_+}.
\end{equation}

Combining all the terms given above, the action becomes,
\beq
{\cal A} &=&  \int_{r_{min}}^{\hat{r}-\epsilon} dr \left[ - L \, y\left(\frac{\lambda \, y^2}{3} + \lambda + R^2\right) + R'\left(R^2+\lambda \right) \log \left|\frac{y+\frac{R'}{L}}{k}\right|\right]
+\int^{\infty}_{\hat{r}+\epsilon} dr \left[- L\,  y\left(\frac{\lambda \, y^2}{3} + \lambda + R^2\right) \right. \\ \nonumber &+&  \left. R'\left(R^2+\lambda \right) \log \left|\frac{y+\frac{R'}{L}}{k}\right|\right] 
-\int dt \, \frac{dR}{dt}\Bigg\{ \left(R^2+\lambda \right) \log \left|\frac{y_{\hat{r}-\epsilon}+\frac{R'_{\hat{r}-\epsilon}}{L}}{k}\right|  - \lambda \, y_{\hat{r}-\epsilon} \frac{R'_{\hat{r}-\epsilon}}{L}\Bigg\}
\\ \nonumber &+&\int dt\, \frac{dR}{dt}\Bigg\{\left(R^2+\lambda \right) \log \left |\frac{y_{\hat{r}+\epsilon}+\frac{R'_{\hat{r}+\epsilon}}{L}}{k}\right|  - \lambda\,  y_{\hat{r}+\epsilon} \frac{R'_{\hat{r}+\epsilon}}{L}\Bigg\}
+\int dt\, \int_{r+\epsilon}^{\infty} \frac{L \, y}{y^2-\left(\frac{R'}{L}\right)^2} \dot{{M_+}} -\int dt {M_+}.
\eeq

This {\em reduced} action still has considerable gauge redundancy
coming from coordinate transformations. We may use this to our
advantage and simplify terms by choosing a particular form for the
fields $R, L$ - however, we have to ensure that the jump constraint
coming from ${\mathcal H}_t=0$ is respected. Put in another way, not
all coordinate transformations are allowed by the constraints -
especially in the region around the shell.

A natural choice of coordinates that fixes this gauge freedom is $R=r$
qand $L=1$. However, if we assume it for $r>(\hat r +\epsilon)$,
because of ${\mathcal H}_t=0$, we cannot assume $R=r$ in a region $r_<
< r< \hat r-\e$ inside of the shell.  In this region the field $R'$ is
not an independent variable - i.e., it is determined by the jump
discontinuity ${\mathcal H}_t=0$ in terms of the variables in the
region $r>\hat r+\epsilon$.

Keeping this in mind, we shall calculate the time derivative of the
above action to obtain the Lagrangian for this system. The derivative
w.r.t $L$ gives $\left( \partial {\cal A}/ \partial L\right) \left(d
L/ dt\right) = \pi_L \dot{L}$.  The contribution from the derivative
w.r.t $R$ involves three terms
\begin{equation}
    \frac{\partial {\mathcal A}}{\partial R} \frac{d R}{dt} 
    + \frac{\partial {\cal A}}{\partial R'} \frac{d R'}{d t}
    = \frac{\partial {\mathcal A}}{\partial R} \frac{d R}{dt} 
  - \frac{\partial}{\partial r}\left( \frac{\partial {\mathcal
      A}}{\partial R'}\right)\dot{R}+
  \frac{\partial}{\partial r}\left(\frac{\partial {\cal A}}
       {\partial R'}\dot{R}\right).
  \end{equation}
The first two terms combine to give $\, \pi_R \dot{R}$, but the second
term involves varying $R'$ over all spacetime - as we have argued, due
to the constraint ${\mathcal{H}}_t=0$, in the range $r_< <r<\hat
r-\e$, we are not allowed to vary $R'$. Thus we subtract the integral
of this term over this range. By an explicit computation, we find that
only the term involving the second derivative of $R$ contributes in
this range \cite{Gooding:2014uqa} which is
\begin{equation}
  \int _{r<}^{\hat{r} - \epsilon}  \Bigg\{\frac{R''}{y L}\Bigg[\lambda
    \Big[1 - \left(\frac{R'}{L}\right)^2\Big]+R^2 -\lambda y^2\Bigg\}\dot{R}.
\end{equation}
In this region, we may as well assume that $y,L$ are essentially
constant, and $\dot{R}=R' \dot{\hat{r}}$. We can now perform the
integral to get \cite{Gooding:2014uqa},
\begin{equation}
  \int _{r<}^{\hat{r} - \epsilon} dr \frac{\partial}{\partial r} \Bigg\{
  -Ly \dot{\hat{r}}\Bigg[\frac{\lambda}{3}y^2
    + \lambda \Big[1 - \left(\frac{R'}{L}\right)^2\Big] +R^2\Bigg]\Bigg\}.
\end{equation}

The remaining part of the integration is quite straight forward, and
we find that the terms rearrange nicely to give the Lagrangian,
\begin{equation}
{\cal L} = \dot{\hat{r}} L\Big\{ \pi_L(y_<) - \pi_L(y_>) \Big\} 
-\dot{\hat{R}}\left(R^2+\lambda\right)\log\Bigg|\frac{y_{r-\epsilon}
  + \frac{R'_{r-\epsilon}}{L}}{y_{<} + \frac{R'_{<}}{L}}\Bigg|
+\int _{r_{min}}^{r-\epsilon} dr \Bigg[\pi_R \dot{R}+\pi_L \dot{L}\Bigg]
+\int ^{\infty}_{r+\epsilon} dr \Bigg[\pi_R \dot{R}+\pi_L \dot{L}\Bigg]
- M_+. \label{Lagrangian}
\end{equation}

To proceed further and obtain a Lagrangian with a single particle
interpretation, we need to use the jump conditions and determine the
relationship between $y_>$ and $y(\hat r -\e)$ and similarly for $R'$.

\section{The jump discontinuities}\label{sec_jump}
As mentioned earlier, we shall use a coordinate system where the
functions $R$, $L$, $N^t$ and $N^r$ are continuous across the shell
while $R', L', \pi_L,$ and $\pi_R$ are allowed to have finite
discontinuities. The discontinuity in $ \pi_L$ can be calculated by
integrating the constraint equation $\mathcal{H}_r^s +
\mathcal{H}^G_r= 0 $ across the shell, which gives
\begin{equation}
\pi_L(\hat{r} + \epsilon) - \pi_L(\hat{r} - \epsilon) =
-\frac{p}{\hat{L}}. \label{1stdis}
\end{equation}

This equation is exactly same as that of general relativity
\cite{Kraus:1994by}. In case of GR, the second junction condition
which determines the jump discontinuity of $R'$ can also be found by
integrating the ${\mathcal H}_t=0$ equation in a straightforward
manner. But in the case of EGB gravity, the situation is different. To
understand the difficulty, we write the explicit form of the
constraint equation:

 \beq \label{1stconstraint}
0 &=& \mathcal{H}_t^s + \mathcal{H}_t^G \\ \nonumber &=& \left(\sqrt{\left(\frac{p}{\hat{L}}\right)^2 + m^2}\right)\,\delta(r-\hat{r}) + y\Bigg\{\pi_R + y\Big[LR - \lambda \left(\frac{R'}{L}\right)'\Big]\Bigg\} - LR \Big[ 1-  \left(\frac{R'}{L}\right)^2\Big]
 + \left(\frac{R'}{L}\right)'\Bigg\{R^2 + \lambda \Big[ 1-  \left(\frac{R'}{L}\right)^2\Big] \Bigg\}.
 \eeq
 
All the terms can be easily integrated across the shell except the
term $ \lambda\, y^2 \left(R' / L\right)'$ since both $y$ and $ R'$
have jump discontinuities at the location of the shell. Unless we know
the {\em exact} dependence of $y$ in terms of other quantities, the
second jump condition cannot be determined.

One way is to solve Eq.(\ref{pilandy}) and express $y$ in terms of
$\pi_L$ which can always be done at least in a series expansion. We
can then substitute the result into the integral and determine the
jump condition (order by order). Such a procedure although possible
in principle is non-trivial to implement because of the
presence of several branches of the solution.

To solve this problem, we instead use a physically motivated condition. First of all, we note that in GR, there is no such problematic term and the jump condition is (Eqn. 3.12 of \cite{Kraus:1994by}, generalized to five dimensions),
\bea
R'(\hat{r}+\epsilon) - R'(\hat{r}-\epsilon) =  - \frac{\sqrt{ p^2 + m^2 \hat{L}^2}}{\hat{R^2}}. \label{2nddis}
\eea
Also, in case of GR, we have $ y = - \left( \pi_L / R^2 \right)$. Let us consider the case of a massless shell and define $\eta = sgn(p) = \pm$. Then the jump condition maybe rewritten as, 
\bea
\left(\frac{R'}{L} +\eta \, y \right) _{\hat{r}+\epsilon} = \left(\frac{R'}{L} +\eta \, y \right) _{\hat{r} - \epsilon}.
\eea

To understand the implication of the continuity of the quantity $
\eta\,y + (R' / L)$ across the shell, we consider the quantity $ d
\hat{R} / dt$ which represents the velocity of the shell, in our
choice of gauge $R=r$. We demand that the shell is modeled such a way
that this velocity of the shell is constant across the shell i.e, we
are considering a structureless shell without any internal
stresses. We will show that this implies the continuity of the
quantity $ \eta \, y + ( R' / L)$ \cite{Fiamberti}.

So, we restrict ourselves to the solutions which satisfy the condition,
\begin{equation}\label{condition}
\frac{dR}{dt} (\hat{r}+\epsilon)= \frac{dR}{dt} (\hat{r} - \epsilon).
\end{equation}
This immediately implies,
\bea
\dot{R}(\hat{r}+\epsilon)+ R'(\hat{r}+\epsilon)\dot{\hat{r}} = \dot{R}(\hat{r}-\epsilon)+R'(\hat{r}-\epsilon)\dot{\hat{r}}.
\eea

We write this as $\dot{\hat{r}} \Delta R'+ \Delta \dot{R} = 0$ where
$\Delta$ represents the jump across the shell. We can also replace the
jump in $\dot{R}$ in terms of the discontinuity of $y$ using the
definition $ y = (\dot{R} - N^rR') / N^t$. Since all the metric
functions like $R$, $L$, $N^t$ and $N^r$ are continuous across the
shell, we obtain,

\begin{equation}\label{trajectory1}
\dot{\hat{r}} = -\frac{\Delta y}{\Delta R'} \hat{N}^t - \hat{N}^r .
\end{equation}

This represents the trajectory of the shell in terms of various jump
conditions. But the trajectory of the shell can also be obtained
directly by varying the action of the shell in the geometry described
by Eq. (\ref{adm}) \cite{Gooding:2014uqa,Fiamberti}. In the massless
limit, the variation gives,

\begin{equation}\label{trajectory2}
\dot{\hat{r}} =  \frac{\hat{N^t} }{\eta\,\hat{L}} - \hat{N^r}.
\end{equation}

Comparing Eq. (\ref{trajectory1}) and Eq. (\ref{trajectory2}) we get
the condition,

\begin{equation}\label{jump}
\Delta \left(\frac{R'}{L} +\eta \, y\right) = 0 \implies \left(\frac{R'}{L} +\eta \, y \right) (\hat{r}+\epsilon) = \left(\frac{R'}{L} +\eta \, y \right) (\hat{r} - \epsilon) .
\end{equation}

Therefore, the continuity of the quantity $ \eta\,y + (R' / L)$ is
equivalent to the statement that the thin shell has velocity
$\dot{\hat{r}}$.  In GR, the jump condition is automatically
compatible with this condition. In case of EGB gravity, we will impose
this as an extra condition and this will allow us to derive the second
jump condition (without explicitly determining $y$ in terms of
$\pi_L$).

Using this condition, we can rewrite the offending terms in the
constraint equation as,
\begin{equation}\label{reexpression}
  y^2 \left(\frac{R'}{L}\right)' = \Bigg[\left(\frac{R'}{L}\right)^2
    - 2 \left(\frac{R'}{L}\right)\left(\frac{R'}{L}+ \eta y\right)
    + \left(\frac{R'}{L}+ \eta y\right)^2\Bigg] \left(\frac{R'}{L}\right)'.
\end{equation}

As is evident, if we use the continuity of $ \eta\,y + (R' / L)$, we
can easily integrate this across the shell. In fact, in Appendix
\ref{appendix}, we show how to explicitly integrate this can obtain a
consistent solution of both the constraint equations.\\

So in conclusion, we will finally consider only a massless shell
obeying continuity of $ \eta\,y +  (R' / L)$ across the shell. The
effective trajectory of the shell will be determined subjected to this
condition. In \cite{Fiamberti,Gooding:2014uqa}, such continuity
conditions were shown to lead to consistent equations of motion of the
shell, and in particular, were necessary in order to determine the
equation of motion for the momentum of the shell.

\section{Equation of motion and tunneling Probability}

Inserting the continuity condition Eq. (\ref{jump}) in the massless
limit into the Lagrangian in Eq. (\ref{Lagrangian}), we get
\begin{equation}
{\cal L} = \dot{\hat{r}} L\Big\{ \pi_L(y_<) - \pi_L(y_>) \Big\} 
-\eta \dot{\hat{R}}\left(R^2+\lambda\right)\log\Bigg|\frac{ \frac{R'_{>}}{L}+ \eta y_{>}}{ \frac{R'_{<}}{L}+ \eta y_{<} }\Bigg|
+\int _{r_{min}}^{r-\epsilon} dr \Bigg[\pi_R \dot{R}+\pi_L \dot{L}\Bigg]+\int ^{\infty}_{r+\epsilon} dr \Bigg[\pi_R \dot{R}+\pi_L \dot{L}\Bigg] - M_+.
\end{equation}
If we set $R=r$ and $L=1$ as a gauge choice in the Lagrangian, the
time derivative terms drop out, and we obtain the canonical momentum
of the particle as,
\begin{equation}
p_c = \pi_L(y_<) - \pi_L(y_>)  - \eta \left(r^2+\lambda\right)\log\Bigg|\frac{ 1+ \eta\, y_{>}}{1+ \eta\, y_{<} }\Bigg|.
\end{equation}
As in the case of Einstein gravity (\cite{Kraus:1994by}), we find that
${M_+}$ is the Hamiltonian of the shell (since the Lagrangian has the
structure of $L=p_c \dot{\hat{r}}-H$).  It is also possible to obtain
an explicit expression for ${M_+}$. \\
  
The trajectory of the massless shell is obtained from the effective
Lagrangian as,

\bea
\frac{ 1}{ \dot{\hat r}} = \frac{ \partial \,p_c}{\partial M_{+}} = \frac{1}{y+\eta} + \frac{\eta \lambda \left(y-\eta\right)}{y\left(\lambda y^2 + r^2\right)} .\label{trajec}
\eea

This represents the effective trajectory of a massless shell moving in
a spherically symmetric black hole spacetime which is a solution of
Einstein-Gauss-Bonnet gravity.  For definiteness, let us now choose
the case $\eta = +1$, corresponding to an outgoing shell. Also, for
our choice of gauge, $ y = - N^{r} / N^{t}$ and solving the ADM mass
equation Eq. (\ref{constraint}) gives $ y = \pm \sqrt{1- f(r)}$ where 
\bea
f(r)=1+\frac{r^2}{\lambda}\Big[1-\sqrt{1+\frac{4\, \lambda \, M_{+}}{r^4}}\Big]. 
\eea
$M_+$ is a function of the momentum $p$ of the shell which may be
determined by using the jump conditions as detailed in the
appendix. But we shall not require the explicit solution in what
follows.

We fix all the signs using the case of general relativity. In general
relativity, (when $\lambda = 0$) the effective trajectory of the shell
is a null geodesic in a Gullstrand-Painlev\'{e} form of the metric
given by \cite{Kraus:1994by},

\bea
ds^2 = - f (r) \,dt^2 + 2 \, \sqrt{ 1- f(r)} \, dt \,dr + dr^2 + r^2 d\Omega^2,
\eea
thus fixing the sign $ y = -\sqrt{ 1- f}$. The back reaction effect
due to the shell of energy $E$ shifts the mass of the original
solution from $M$ to $M_{+} = M - E $. In the presence of the
Gauss-Bonnet term, the trajectory Eq. (\ref{trajec}) of the shell is
modified and the shell no longer moves along a null geodesic of the
original metric. \\

Once we obtain the trajectory of the shell, we can proceed to
calculate the quantum mechanical emission probability. In
\cite{Kraus:1994by}, this is determined by quantizing the effective
Lagrangian of the shell and finding the Bogoliubov coefficients using
a WKB approximation. In principle, we can repeat the same calculation
with the trajectory in Eq.  (\ref{trajec}). But, instead, we shall
follow an equivalent but more physically transparent procedure using
the results of \cite{Parikh:1999mf, Sarkar:2007sx}. The key idea is to
calculate the tunneling probability of such a shell moving in the
geometry corrected by the back reaction by evaluating the imaginary
part of the action due to the presence of the horizon. \\

We begin with the action and find the imaginary part of the action due
to the presence of the horizon at $ f ( r, M_{+} ) = 0$. The
semi-classical action for an s-wave outgoing positive energy shell
which crosses the horizon outwards from $r_h - \epsilon$ to $r_h +
\epsilon$ is,

 \beq\label{GBBG}
{\cal A } = \int p\, dr = \int \int_{0}^{p} dp' dr = \int_{r_h + \epsilon}^{r_h - \epsilon}dr \int_{0}^{p} \frac{d {\bf M}}{\dot{ \hat r}} .
\eeq

Where we have used the Hamiltonian equation $ \left( d {\bf M} / d p
\right) = \dot{\hat r}$ and $d{\bf M} $ is the change in ADM
Hamiltonian due to the tunneling of the shell, so in our case $d {\bf
  M} = M - M_{+}$, the shift in the ADM mass of the space time due to
back reaction of the shell.\\

The crucial step is then to consider the trajectory of the shell near
the horizon. The presence of the horizon leads to a pole in the
semi-classical action and when extended to the complex plane, the pole
gives rise a residue contributing to the imaginary part of the
action. Therefore, we only need to consider the trajectory in a near
horizon approximation. In terms of the function $f(r)$, the trajectory
of an outgoing shell is,
\begin{equation}\label{GBBF}
  \frac{1}{\dot{\hat{r}}} = \frac{1}{1-\sqrt{1-f}}
  + \frac{\lambda (1+\sqrt{1-f})}{\sqrt{1-f}\left(\lambda(1-f)+r^2\right)}.
\end{equation}
The second term of the trajectory in Eq. (\ref{GBBF}) is finite. The
contribution to the imaginary part is entirely due to the first term
and near the horizon at $ r = r_h$ Eq. (\ref{GBBG}) can be written as,

\beq
    {\cal A} = \int \int_{r_h + \epsilon}^{r_h - \epsilon}
    \frac{d {\bf M} \,dr}{\kappa\, (r - r_h)},
\eeq
where we have defined the surface gravity as $\kappa = f'(r_h) / 2$. To 
evaluate the imaginary part, we evaluate the residue at this pole,
\beq
    {\cal A} =  ( - i \pi)\int \frac{{d \bf M}}{ \kappa}
    = (- i)\int \frac{{\pi\,d \bf M}}{\kappa}.
\eeq
Then, the total semi-classical tunneling probability is,
\bea
\Gamma   \propto e^{- 2\, {\bf Im} {\cal A}} \propto
e^{\int \frac{{ 2 \pi\,d \bf M}}{\kappa}}.
\eea
Using the first law for black holes in EGB gravity expressed in
Eq. (\ref{1stlaw}), we can express this tunneling probability as,

\bea
\Gamma \propto e^{ \int \delta S} = e^{\Delta S},
\eea
where $\Delta S$ is the change in the entropy as the ADM mass changes
from $M$ to $M_{+} = M - E$, due to the tunneling of a shell of energy
$E$. Note that, here the entropy is no longer proportional to the area
but given by the Jacobson-Myers expression in Eq. (\ref{JMent}). \\

In conclusion, the effective trajectory of the shell in EBG gravity is
no longer a null geodesic in the corresponding Gullstrand-Painlev\'{e}
form of the metric. However, the pole structure due to the presence of
the horizon remains same as in the case of general relativity and a
calculation based on semi-classical tunneling gives the emission
probability of a shell of energy $E$ as,
\be
\Gamma(E) \propto e^{\Delta S},
\ee
where $\Delta S = S(M) - S(M-E)$, the change in entropy because of the
change in the ADM mass due to the back reaction of the shell. Note
that, the entropy $S$ is exactly same as the expression which obeys
the first law in Eq. (\ref{1stlaw}). Both the background field
equation, as well as the entropy pick up corrections due to the
Gauss-Bonnet term, but the tunneling probability continues to be the
same form as in the case of general relativity. \\

As discussed in the introduction, as long as the tunneling probability
is given by the exponential of the change of the entropy, there will
be no correlation between the emission of two successive shells since
$ \Gamma (E_1 + E_2) = \Gamma(E_1) + \Gamma(E_2)$. Hence, the
correction due to the Gauss-Bonnet term does not create any new
correlation in the semi-classical Hawking spectrum and therefore any
hope of solving the information loss paradox using modified dynamics
of gravity may not be fruitful. As long as we are in the regime of the
validity of semiclassical gravity, it seems that information loss is
inevitable.

\section{Discussion}

The inclusion of higher curvature terms in the action changes the
dynamics of gravity. Therefore, it seemed plausible that the
self-interaction calculations in \cite{Kraus:1994by} could be modified
significantly in the presence of higher curvature terms and lead to
new correlations in the Hawking spectrum. This possibility motivated
us to generalize the results of \cite{Kraus:1994by} beyond general
relativity. We consider an Einstein-Gauss-Bonnet theory in five
dimensions and calculate the effective trajectory of a massless shell.
The procedure is almost parallel to that in the case of general
relativity, except that the jump conditions were handled differently.
We imposed a continuity condition on the velocity of the shell which
was automatic in GR (see the discussion in
\cite{Fiamberti,Gooding:2014uqa}) and this enabled us to integrate the
jump condition across the shell. \\

The characteristic feature of the Gauss-Bonnet terms was that the
constraint structure is same as general relativity. There are no
additional degrees of freedom and therefore the calculations turned
out to be straightforward. The effective equation of a massless shell
can be obtained and is given by Eq. (\ref{trajec}). The trajectory is
used to calculate the semi-classical tunneling probability and the
result is the same form as of general relativity $\Gamma (E) \propto
e^{\Delta S}$ where the entropy $S$ is now the Jacobson Myers entropy
in Eq. (\ref{JMent}) obeying the first law. Also, as in case of GR, $
\Gamma (E_1 + E_2) = \Gamma(E_1) + \Gamma(E_2)$ implying that there is
no correlation between successive emission of two shells.\\

Although, it has been argued that small corrections in Hawking
spectrum can not lead to the purification of the final state
\cite{Mathur:2009hf, Mathur:2011wg}, it is interesting to note that a
spherically symmetric self-interaction calculation does not produce
any useful correlation at all, even when the dynamics of gravity is
modified by the Gauss-Bonnet term.  Clearly, it will
be worthwhile to extend our result beyond Lovelock theories and
analyze the situation in the presence of extra degrees of freedom
and/or extra constraints. \\

It has been suggested \cite{Visser:1997yu}, that the original
derivation of black hole radiation may not be sensitive to the details
of the dynamical structure of field equations but rather depends only
on the kinematical properties of the event horizon. Our result is in
conformity with this point of view and seems to suggest that the
proper resolution of information loss is beyond the scope of
semi-classical gravity.

\section{Acknowledgements}

The authors would wish to thank the organizers of The
GIAN workshop at IIT Gandhinagar, where this work was initiated. SS is supported by the Department of Science and Technology,
Government of India under the SERB Fast Track Scheme for Young
Scientists (YSS/2015/001346). KPY likes to thank IIT, Gandhinagar for
hospitality, where part of this work was carried out.

\section{Appendix: A consistency relation for the constraints.}\label{appendix}

In section \ref{sec_jump}, we have shown how the assumption of that
the velocity of the shell is well defined allows us to obtain the jump
conditions for $ \pi_L$ and $ R'$ across the shell (\ref{jump}). In
this appendix, we show that the discontinuities of the field variables
across the shell are all mutually consistent and show that there exist
real solutions such that the constraints are satisfied.

As mentioned earlier, we shall use a coordinate system where the
functions $R$, $L$, $N^t$ and $N^r$ are continuous across the shell
while $R', L', \pi_L,$ and $\pi_R$ are allowed to have finite
discontinuities. For simplicity, we consider the shell is located
between $ r = \hat{r}$ and $ r = \hat{r} +\epsilon$ and we take a
simple choice of the metric function across the shell as
\cite{Gooding:2014uqa},

\begin{equation}
 R(r,t) = r - \epsilon\, \alpha\, g\left(\frac{\hat{r}-r}{\epsilon}\right) ; \,\,\,  L = 1 ,
\end{equation}
where $ g(x)$ is a continuous and differentiable function defined for
all $ x \in (0 , -1)$ and satisfies the following,
 \begin{equation*}
 g (0) = g(-1) = 0 ;\quad g'(0) = 1 ;\quad g'(-1) = 0,
 \end{equation*}
 and time appears implicitly through $\hat{r},\alpha$.
 Differentiating $ R(r,t)$ w.r.t $ r$ gives,
 \begin{equation*}
 R'_{\hat{r}} = 1 + \alpha ;\quad R'_{\hat{r} + \epsilon} = 1.
 \end{equation*}
The variable $y$ is now determined so that the velocity continuity condition
(\ref{jump}) is satisfied, 
\begin{equation*}
y_{\hat{r}+\epsilon} = h ;\quad y_{\hat{r}} = h - \eta \,\alpha ;\qquad \eta = sign(p).
\end{equation*}
The value of $ h(t)$ can be readily obtained from the ADM condition Eq.(\ref{constraint}) as,
\bea \label{hsolution}
h = \pm \sqrt{1- f(r,M_+)} \,;\quad f(r,M_+)=1+\frac{r^2}{\lambda}\Big[1-\sqrt{1+\frac{4\, \lambda \, M_{+}}{r^4}}\Big]. 
\eea

Now the discontinuity of $ \pi_L$ (\ref{1stdis}) can be written as,
\begin{equation}
\frac{2\, \eta \, \lambda \, \alpha^3}{3} - \eta\, \lambda\, h^2 \,\alpha - 2 \lambda \,h \,\alpha - \eta \,\alpha\, X^2 + 2\, \eta\, \lambda\, \alpha^2 = -p.
\end{equation}
The second constraint relation is obtained by integrating $
\mathcal{H}_t^s + \mathcal{H}_t^G = 0 $ across the shell
(\ref{1stconstraint}). We assume that the momentum $\pi_R$ and $y$ are
both non-singular at the location of the shell. Using the velocity
continuity condition as in Eq.(\ref{reexpression}),
%% This implies that the only terms that contribute to the
%% discontinuity are $$\left(\sqrt{\left(\frac{p}{\hat{L}}\right)^2 +
%% m^2}\right)\,\delta(r-\hat{r}) - \lambda\,y^2
%% \left(\frac{R'}{L}\right)' + \lambda
%% \left(\frac{R'}{L}\right)'\Big[ 1- \left(\frac{R'}{L}\right)^2\Big]
%% + \left(\frac{R^2R'}{L}\right)'$$ which we can rewrite using the
%% velocity continuity condition Eq.(\ref{reexpression})
%% as $$\left(\sqrt{\left(\frac{p}{\hat{L}}\right)^2 +
%% m^2}\right)\,\delta(r-\hat{r}) +\Bigg\{ -
%% \frac{2\lambda}{3}\left(\frac{R'}{L}\right)^3 + \lambda
%% \left(\frac{R'}{L}\right)+ \left(\frac{R^2R'}{L}\right)\Bigg\}'$$
we may integrate across the shell and then substitute to get
\begin{equation}
\frac{2\, \lambda\, \alpha^3}{3} + \lambda\, \alpha^2 - \eta\, \lambda\, \alpha^2 h + \lambda\, h^2 \,\alpha - \alpha \,X^2 = -\hat{V}.
\end{equation}
These two equations show that the values of the fields $R', y$ inside
the shell are fixed by their values outside the shell.

Solving the above two equations in the massless limit leads to the relation
\begin{equation}
2\, h^2 + \eta\, h(2-\alpha) -\alpha= 0.
\end{equation}

This equation can be solved for $ \alpha(t)$ in terms of $ h(t)$
which, in turn, is determined as a solution of either cubic equations
in terms of the $\hat r$ variables of the shell.  Then $M_+$ is
determined in by $h = \pm \sqrt{1- f(r,M_+)}$

\end{document}